\documentclass{ws06-procs}

\begin{document}

\title{Cosmological Voids: Observational Evidence and Models}

\author{M. Serpico}

\address{Department of Physical Sciences, University ``Federico II'', via Cinthia, 80126 Napoli, Italy\\E-mail: serpicom@na.infn.it}

\maketitle

\abstracts{Cosmological Voids are among the largest structures known
in the universe with diameters ranging from 20 Mpc to 80 Mpc. In
this paper I present a short review of the observational evidence
and models so far proposed and a discussion of possible
observational tests for these models.}

\section{Observational Evidence}
The first observation of a large region of universe not containing
any visible matter was made in 1981 by Kirshner et
al.\cite{kirshner}. Starting from the discovery of this huge void
(about $80 Mpc$ in diameter) of matter in Bo\"{o}tes, galaxy surveys
analysis showed that a considerable fraction of the volume of the
universe is occupied by empty regions containing only few visible
galaxies. These regions have been named Cosmological Voids. Studying
the characteristic features of these structures is surely relevant
to a deeper understanding of the large scale structure of the
universe. Nevertheless, present observational data allow to outline
some general features only: cosmological voids seem to occupy about
$40\%$ in volume of the universe; they have quasi-spherical
simmetry; they have typical diameters of the order of $10\div 20
Mpc$; the morphological properties of galaxies located inside or
nearby a void are not considerably different from the cosmological
background.

The most detailed observational studies on the subject have been
made by a research group working in USA at the Departement of
Physics of Drexel University in Philadelphia (see, for example,
Hoyle et al.\cite{hoyle}). These studies are based on an automated
alghorithm developed by El-Ad et al.\cite{el-ad} and named
\emph{Voidfinder}, which allows to construct a void survey starting
from a galaxy survey.

Let's briefly outline how \emph{Voidfinder} works. The analysis of
the galaxy survey starts by dividing the galaxies in two subgroups:
void galaxies and boundary galaxies. Void galaxies are those having
not more than two neighbour galaxies in a sphere of radius \emph{r}
centered on them (usually \emph{r} is fixed to $5 Mpc$). All the
remaining galaxies in the survey are boundary galaxies. The
algorithm then divides the volume occupied by the galaxy survey in
cubical cells of appropriate volume (chosen by user) and search for
those cells not containing boundary galaxies. Now an approximated
procedure determines the positions of the centers and radii of
comoving spheres containing no boundary galaxies. Those are the
cosmological voids contained in the galaxy survey.

Applying \emph{Voidfinder} to the 2dFGRS galaxy survey, Hoyle et al.
find 289 cosmological voids with average radius of $(12\pm2) Mpc$,
density contrast with respect to the cosmological background of
$-0.95$ and occupying about $40 \%$ of the volume of the survey.

\section{Models}

The observational evidences mentioned in the previous section has
generated great interest concerning the construction of models
capable of explaining the actual distribution of matter inside
cosmological voids and their formation mechanism.

It is generally accepted that there is strong correlation between
the morphology of galaxies and the density of their formation
environment. More in particular, elliptical galaxies are more likely
to form in higher density environments (for example in clusters)
while spirals are more common where the density is lower. This
morphology-density correlation, originally foreseen in the so called
\emph{biased galaxy formation picture}, is supported by several
observational studies (see for example Goto et al.\cite{gotoetal})
and is particularly relevant to investigate the possible presence of
matter inside the voids. In fact, if the voids are completely
devoided of matter as suggested by observations, then the galaxies
we observe nearby their boundaries should present a strongly
characteristic morphological distribution because they formed in an
enviroment in which the density of matter is rapidly falling from
values near to the average cosmological background (outside the
void) to extremely low values (inside the void). So far there is no
conclusive observational evidence concerning the existence of this
characteristic distribution. To solve this apparent contradiction,
we should admit the possibility that the internal part of the voids
contains some form of dark matter (as originally stated by
Peebles\cite{peebles}). Already in Kirshner et al.\cite{kirshner},
the authors stated that inside the void could be present an
over-density of galaxies with very low brightness that we are still
not able to observe. This is the main open issue related to
cosmological voids.

Another important issue is the one concerning the models for void
formation. Among the various models presented so far, we may quote
the one originally introduced by Friedmann and
Piran\cite{Friedmann}, accordingly to which voids form from the
comoving expansion of negative primordial perturbations in the
density field. Several N-body simulations of this formation
mechanism based on the cold dark matter scenario produced results
which are consistent with observational data.

More recently another formation mechanism was proposed by
Stornaiolo\cite{Stornaiolo}. In this scenario the collapse of
extremely large wavelength\footnote{The wavelenght of initial
perturbations is assumed to be equal to void present diameters.}
positive perturbations led to the formation of low density/high mass
black holes (Cosmological Black Holes or CBH). Voids are then formed
by the comoving expansion of the matter surrounding the collapsed
perturbation. According to this model, at the center of each
cosmological void (assumed to be spherical) a CBH having a mass
\begin{equation}\label{Voidmass}
M_{void}=\frac{4}{3}\pi\Omega_{cbh}\rho_{c0}R^{3}
\end{equation}
is present. In (\ref{Voidmass}) the parameter
\begin{equation}\label{Omegacbh}
\Omega_{cbh}=\frac{\rho_{cbh}}{\rho_{c0}}
\end{equation}
represents the ratio of the density $\rho_{cbh}$ of all the CBHs
contained in the universe to the critical density as observed at the
present time $\rho_{c0}$ and $R$ is the radius of the void. The mass
of the CBH partially or completely ``compensate'' the lack of matter
in the volume occupied by the void according to the values of
$\Omega_{cbh}$. The actual value of this parameter has to be
adjusted according to future observational estimates of CBH mass.
Nevertheless, at this stage, it is reasonable to assume that
$\rho_{cbh}$ has the same value of the observed matter density
$\rho_{matter}$ scaled by the fraction of the volume of the universe
occupied by the voids. This assumption implies
$\Omega_{cbh}\simeq0.2$. In a void about $25 Mpc$ in diameter we
then have a CBH with mass of the order of $10^{14} M_{\odot}$.

\section{Testing the Models. Voids as Gravitational Lenses}
Observational studies concerning cosmological voids are mainly
intended to investigate their matter content. So this section will
be focused on this issue only. To understand if there is matter
inside a void one can use essentially three main kinds of
observational studies: studying the morphology of galaxies nearby
the void; studying the dispersion of the peculiar velocities of
galaxies nearby the void; studying the void as a gravitational lens
deflecting the light coming from background sources and/or producing
perturbations on the CMB anisotropy.

I have already discussed the morphology of ``void galaxies'' in the
previous section. What can be said about the dispersion of
velocities is that, as in the case of galaxy morphology, the
observations made so far suggest that there could be some ammount of
matter inside the voids\cite{peebles} but they are not conclusive.

The gravitational lensing properties of voids have so far remained a
topic treated only in theoretical papers. This is due to the fact
that if the voids are filled with low brightness (and, because of
this, not visible) ordinary galaxies then they would have no
observable peculiar gravitational lensing effect and, as pointed out
by Amendola et al.\cite{amendola}, this is true even in case the
voids are completely devoided of matter. One case of interest is
represented by the model proposed by Stornaiolo\cite{Stornaiolo}. In
the article\cite{serpico}, the coauthors and me have presented the
results of several simulations of gravitational lensing effects
produced by a cosmological void structured according to this model.
With regard to its lensing properties, the CBH contained in the void
acts on background sources as a Schwarzschild lens. According to our
simulations the CBH should produce three potentially observable
lensing effects which are described in the following.

A Schwarzschild lens produces double images of background sources
having a caratheristic angular separation which depends on the mass
and distance of the lens. For ordinary Swarzschild lenses this
angular separation assumes typical values of about 10 seconds of
arc. A typical CBH located at comoving distance of $80 Mpc$ from us
and with a mass of $4\times 10^{14} M_{\odot}$ produces several
double images with angular separation of about 6 primes of arc. Once
one knows the position of the center of a void (and its distance and
radius) in a galaxy survey, he can search for couples of galaxies
nearby this center and having angular separation of about 6 primes
of arc. Comparing spectra one can then verify the presence of double
images.

Each double image produced by a Schwarzschild lens is composed by a
quite strongly magnified primary and by a fainter secondary image.
Due to the production of many secondary images of background
galaxies a typical CBH considerably alters the number of faint
galaxies with redshifts larger than that of the center of the void
and observed in its proximity. This effect could be observed
performing a galaxy count in a suitable solid angle centered on the
void and verifying the presence of an increase in the number of
faint ($m>21$) galaxies with respect to galaxy number counts
performed in an equal solid angle not containing any void.

The production of double images together with the magnification
effect also determine a characteristic profile in the observed
radial distribution of galaxies around the CBH. This is a peculiar
feature of the CBH lensing effects, which is due to its huge mass
and would not be present for any ordinary Schwarzschild lens. For a
detailed description of this effect see Serpico et
al.\cite{serpico}.

The three gravitational lensing effects I have mentioned are all
strong signatures of the existence of Cosmological Black Holes. The
possibility to observe these effects is connected to the use of a
galaxy survey deep enough in magnitudes and to the necessity of
having a suitable void survey derived from the same galaxy survey.
We hope the Sloan Digital Sky Survey could serve to this scope in
the future and that the tests we proposed for the CBH-void model
could be conclusive. In the future we are also planning to
investigate the CBH effect on the Cosmic Microwave Background.

\end{document}